\documentclass[12pt,preprint]{aastex}

\shorttitle{Flux rope} \shortauthors{Li & Zhang}

\begin{document}

\title{Fine-scale Structures of Flux Ropes Tracked by Erupting Material}

\author{Ting Li\altaffilmark{1} \& Jun Zhang\altaffilmark{1}}

\altaffiltext{1}{Key Laboratory of Solar Activity, National
Astronomical Observatories, Chinese Academy of Sciences, Beijing
100012, China; [liting;zjun]@nao.cas.cn}

\begin{abstract}

We present the \emph{Solar Dynamics Observatory} observations of two
flux ropes respectively tracked out by material from a surge and a
failed filament eruption on 2012 July 29 and August 04. For the
first event, the interaction between the erupting surge and a
loop-shaped filament in the east seems to ``peel off" the filament
and add bright mass into the flux rope body. The second event is
associated with a C-class flare that occurs several minutes before
the filament activation. The two flux ropes are respectively
composed of 85$\pm$12 and 102$\pm$15 fine-scale structures, with an
average width of about 1$\arcsec$.6. Our observations show that two
extreme ends of the flux rope are rooted in the opposite polarity
fields and each end is composed of multiple footpoints (FPs) of the
fine-scale structures. The FPs of the fine-scale structures are
located at network magnetic fields, with magnetic fluxes from
5.6$\times$10$^{18}$ Mx to 8.6$\times$10$^{19}$ Mx. Moreover, almost
half of the FPs show converging motion of smaller magnetic
structures over 10 hr before the appearance of the flux rope. By
calculating the magnetic fields of the FPs, we deduce that the two
flux ropes occupy at least 4.3$\times$10$^{20}$ Mx and
7.6$\times$10$^{20}$ Mx magnetic fluxes, respectively.

\end{abstract}

\keywords{Sun: corona --- Sun: filaments, prominences --- Sun:
coronal mass ejections (CMEs)}

\section{Introduction}

The flux rope is thought to be closely related to the coronal mass
ejection (CME), which generally has a three$-$part structure: the
bright core, the dark cavity and the leading edge (see e.g., Illing
\& Hundhausen 1986). It is often believed that the twisted flux rope
is the dark cavity which accumulates magnetic energy and mass within
it (Chen 1996; Hudson \& Schwenn 2000; Gibson et al. 2006). The
ideal magnetohydrodynamic (MHD) instabilities of the flux rope such
as the kink instability and the torus instability are thought to be
one type of mechanism that triggers the CME and associated
activities (T{\"o}r{\"o}k \& Kliem 2003, 2005; Fan 2005; Kliem \&
T{\"o}r{\"o}k 2006; Olmedo \& Zhang 2010). Thus, a detailed study of
the flux rope is important for a clear understanding of CMEs, and
this leads to a good ability to forecast CMEs and associated space
weather.

The flux rope has been reconstructed in a complex magnetic topology
from observed vector magnetograms by using non linear force-free
field models (Guo et al. 2010; Canou \& Amari 2010; Jing et al.
2010). Moreover, the formation and dynamic behavior of flux ropes
have been simulated successfully by many authors (Amari \& Luciani
1999; Amari et al. 2010; Fan \& Gibson 2004). In the simulation of
Aulanier et al. (2010), the flux rope is progressively formed by
flux-cancellation-driven photospheric reconnection in a bald-patch
separatrix.

Recently, the direct observations of flux ropes during the eruption
process have been reported by using the data from the Atmospheric
Imaging Assembly (AIA; Lemen et al. 2012) onboard the \emph{Solar
Dynamics Observatory} (\emph{SDO}; Pesnell et al. 2012). Cheng et
al. (2011) presented the observation of a flux rope as a bright blob
of hot plasma in the channel of 131 {\AA}. The hot flux rope rapidly
moved outward and stretched the surrounding magnetic field upward,
similar to the classical magnetic reconnection scenario in eruptive
flares. Detailed thermal property of the flux rope based on
differential emission measure was investigated by Cheng et al.
(2012). However, there is to our knowledge no observations of
fine-scale structures of flux ropes tracked by erupting material. In
this letter, we present two events of flux ropes and investigate
their fine-scale structures and magnetic properties by using the
data from \emph{SDO}/AIA and the Helioseismic and Magnetic Imager
(HMI; Schou \& Larson 2011).

\section{Observations and Data Analysis}

On 2012 July 29 and August 04, \emph{SDO}/AIA observed two flux
ropes that are respectively tracked by material from a surge and a
failed filament eruption. Both of them are composed of thread-like
structures, which warp and interweave together.

The \emph{SDO}/AIA takes full-disk images in 10 (E)UV channels at
1$\arcsec$.5 resolution and high cadence of 12 s. The flux ropes
could be observed in all the 7 EUV channels. The 171 channel best
shows the flux rope and we focus on this channel in this study. We
also present the observations of flux ropes in different channels
such as 304, 193, 335 and 131 {\AA}.

The 5 EUV channels correspond to different temperatures: 171 {\AA}
(Fe IX) at 0.6 MK, 304 {\AA} (He II) at 0.05 MK, 193 {\AA} (Fe XII)
at 1.5 MK (with a hot contribution of Fe XXIV at 20 MK and cooler O
V at 0.2 MK), 335 {\AA} (Fe XVI) at 2.5 MK and 131 {\AA} (Fe VIII,
Fe XXI) at 11 MK (O'Dwyer et al. 2010; Boerner et al. 2012; Parenti
et al. 2012). We also use the full-disk line-of-sight magnetic field
data from the HMI onboard \emph{SDO}, with a cadence of $\sim$ 45 s
and a sampling of 0$\arcsec$.5 pixel$^{-1}$.

\section{Results}

\subsection{Overview of the Two Flux Ropes}

At about 00:30 UT on 2012 July 29, a surge occurred in NOAA AR 11530
(S19W00) and plenty of material was ejected northward (see Figure
1\emph{b} and Animation 1 in the online journal). By examining the
\emph{SDO}/HMI line-of-sight magnetograms, we found that the
magnetic flux cancellation took place several hours prior to the
surge (Figure 1\emph{d}). The obvious brightening was observed at
171 {\AA} at the location of the cancelled flux. This cancellation
could be what led to the occurrence of the surge. When the surge
first appears and starts to ascend, there is evidence of interaction
between it and a loop-shaped filament in the east side of the
erupting surge (see Figure 1\emph{e} and Animation 2 in the online
journal). This interaction seems to ``peel off" the filament and to
add mass into the flux rope body. Simultaneously, brightenings at
the interaction location and the footpoint of the surge at 304 {\AA}
are also observed (see Figure 1\emph{e} and Animation 2 in the
online journal).

At about 00:52 UT, the erupting material seemed to be confined and
moved toward the west (see Animation 1 in the online journal).
Meanwhile, the bright fine-scale structures are clearly observed.
About 52 min later (01:44 UT), the erupting material arrived at the
west end of the flux rope and the entire flux rope was tracked out
by the ejected material (see Figure 1\emph{c}). It seems that the
surge occurs within the flux rope and the footpoint of the surge is
located at one end of the flux rope, and thus the material from the
surge flows along the flux rope. The approximate length of the flux
rope is 596 Mm, and the apparent flow speed of the material along
the flux rope body is about 150 km s$^{-1}$. The flux ropes observed
at 304, 193, 335 and 131 {\AA} are roughly the same on the whole.
However, they are different in some details. Taking the area denoted
by white rectangles in Figure 1 for example, the 193 {\AA}
observations are similar to those of 171 {\AA}, and the 304, 335 and
131 {\AA} observations are different from 171 {\AA} observations as
the fine-scale structures (pointed by white arrows) are not clearly
identified in these three channels.

On 2012 August 04, the second flux rope was observed in NOAA AR
11539 (S23E32). Before the flux rope was tracked, there existed a
filament at the east part of the flux rope (see Figure 2\emph{a}).
At 11:04 UT, a C2.9 flare occurred at the east of the filament (see
Figure 2\emph{e} and Animation 4 in the online journal), which
peaked at 11:47 UT and ended at 12:49 UT. At about 11:14 UT, the
filament started to turn over and brighten (Figure 2\emph{b}). At
about 11:40 UT, the ejected material from the filament successively
moved toward the southwest (see Animation 3 in the online journal).
Then the arch-shaped flux rope with helical fine-scale structures
was observed clearly (Figure 2\emph{c}). The observed length of the
flux rope is about 546 Mm, and it takes the filament material about
51 min to flow from the east to the west end. The apparent flow
velocity along the flux rope body is approximately 180 km s$^{-1}$.
Similar to the first flux rope on 2012 July 29, the fine-scale
structures denoted by the white arrow in white rectangles of Figure
2 are identified clearly at 171 and 193 {\AA} and seems obscure at
304, 335 and 131 {\AA}. At about 16:00 UT on August 6, the east part
of the second flux rope erupted, and this eruption resulted in a
faint CME with a speed of about 260 km s$^{-1}$. The whole flux rope
erupted at about 03:00 UT on August 8, associated with a halo CME
with a speed of about 230 km s$^{-1}$.

\subsection{Fine-scale Structures of the Two Flux Ropes}

As the erupting material arrived at the west extreme end of the
first flux rope on 2012 July 29, the west end showed obvious
brightening (Figures 1\emph{c} and 3\emph{a}). Then partial material
went back toward the east and brightened the east end at about 03:02
UT (see Figure 3\emph{d} and Animation 1 in the online journal). The
brightening at the ends makes it possible to determine the ends
location. Thus we select two areas (denoted by red and blue
rectangles in Figure 1\emph{c}) to investigate the ends and
fine-scale structures.

By counting the number of fine-scale structures one by one, we
notice that the first flux rope is composed of 85$\pm$12 fine-scale
structures, and 15 well identified fine-scale structures are
selected to measure their widths. Two examples are shown in Figures
3\emph{c}, \emph{f} and \emph{g}. Firstly, the intensity-location
profiles (black curves in Figures 3\emph{f} and \emph{g}) along
slices perpendicular to the fine-scale structures (Slices ``S1" and
``S2" in Figure 3\emph{c}) are obtained. Secondly, we use Gaussian
function to fit the intensity-location profiles and two Gaussian
fitting profiles (blue and red ones) are shown in Figures 3\emph{f}
and \emph{g}. The full width at half maximum (FWHW) of the Gaussian
fitting profile is thought to be the width of fine-scale structure.
The average width of these fine-scale structures is 1$\arcsec$.8,
with the maximum value of 2$\arcsec$.0 and the minimum value of
1$\arcsec$.4.

For the first flux rope, there are 12 western footpoints (FPs) of
the fine-scale structures that form the west end of the flux rope
(white circles in Figure 3\emph{a}). By comparing the 171 {\AA}
observations with line-of-sight magnetograms, we find that all the
western FPs are rooted in negative polarity fields (Figure
3\emph{b}). The net magnetic fluxes of these FPs are in the range of
8.6$\times$10$^{18}$$-$8.6$\times$10$^{19}$ Mx. The magnetic flux of
the west end of the flux rope is $-$4.3$\times$10$^{20}$ Mx. This is
the lower limit of the magnetic flux since some FPs are not
identified and calculated for they are not accompanied by
brightening. The 8 eastern FPs are rooted in positive polarity
fields (Figures 3\emph{d} and \emph{e}). The net magnetic fluxes of
eastern FPs are from 5.6$\times$10$^{18}$ to 2.8$\times$10$^{19}$
Mx. The magnetic flux of the east end of the flux rope is
1.3$\times$10$^{20}$ Mx, which is less than that of the western end.
Not all the eastern FPs of the fine-scale structures are identified
and thus there exists the discrepancy between the two sides.

For the second flux rope on 2012 August 04, we similarly select two
areas where the brightening occurs at the ends (Figures 2\emph{c}
and 4). This arch-shaped flux rope seems more complex than the first
one and has two main western ends (one in Figure 4\emph{a} and the
other one in Figure 4\emph{d}) which are separated apart. The flux
rope is composed of 102$\pm$15 fine-scale structures. The average
width of 20 clearly identified fine-scale structures is
1$\arcsec$.5. The thickest one has a width of 1$\arcsec$.7, and the
thinnest one has that of 1$\arcsec$.1 (Figures 4\emph{c}, \emph{f}
and \emph{g}). The line-of-sight magnetograms show that 22 western
FPs (15 ones in Figures 4\emph{a}$-$\emph{b} and 7 ones in Figures
4\emph{d}$-$\emph{e}) of the fine-scale structures are anchored at
positive polarity fields (Figures 4\emph{a}$-$\emph{b},
\emph{d}$-$\emph{e}). The net magnetic fluxes of these western FPs
are in the range of 1.1$-$8.1$\times$10$^{19}$ Mx. The total
magnetic flux of western ends of the flux rope is
7.6$\times$10$^{20}$ Mx. The eastern end of the flux rope is close
to the solar limb and not accompanied by EUV enhancements, thus it
could not be identified.

By examining the magnetic field evolution at the FPs of the
fine-scale structures over 10 hr before the appearance of the flux
rope, we find almost half of these FPs show converging motion of
smaller magnetic structures for both the flux ropes. Figure 5
presents one example of the eastern FPs of the first flux rope. As
seen in the stack plot along Slice ``A$-$B", the west magnetic
structure obviously moved toward the east one with a velocity of 0.2
km s$^{-1}$ and the motion of the east one was slower than the west
one, with a velocity of 0.1 km s$^{-1}$ (Figure 5\emph{e}).

\section{Summary and Discussion}

We present the \emph{SDO}/AIA observations of two flux ropes on 2012
July 29 and August 04 which are tracked out by material from a surge
and a failed filament eruption. For the two flux ropes, the apparent
speeds of filling of the flux rope structure with chromospheric and
coronal plasma are respectively 150 and 180 km s$^{-1}$, which are
comparable to the typical sound speed for the corona of about
100$-$200 km s$^{-1}$. For the first flux rope, the approximate
length of the flux rope is 596 Mm. By examining the fine-scale
structures which are observed more clearly, we roughly estimate the
twist is about $\pi$. For event 2, the observed length of the flux
rope is about 546 Mm, and the average twist is about 2$\pi$. Seen
from the \emph{Solar-Terrestrial Relations Observatory}
(\emph{STEREO}; Kaiser et al. 2008) B viewpoint, the two flux ropes
are both located at the west limb. By using three-dimensional
reconstructions, we obtain the heights of the two flux ropes, which
are respectively 90 and 140 Mm above the solar surface. The two flux
ropes analyzed here are respectively composed of 85$\pm$12 and
102$\pm$15 fine-scale structures, which probably outline the
magnetic field structures of flux ropes (Martin et al. 2008; Lin
2011). The width of the fine-scale structures ranges from
1$\arcsec$.1 to 2$\arcsec$.0, with an average of about 1$\arcsec$.6.
It is comparable to the resolution limit of the AIA telescope of
about 1$\arcsec$.2, which suggests that even thinner structures may
exist. Moreover, the width of the fine-scale structures is an order
of magnitude larger than the ultrafine magnetic loop structures
observed by Ji et al. (2012).

Before the flux ropes are tracked out by erupting material, part of
magnetic flux rope structures may exist in the space filled in by
the flux ropes. For event 1, there are several similar events in two
days before the appearance of the first flux rope on 2012 July 29.
Moreover, the 304 {\AA} images shortly before the surge injection
and flux rope appearance reveal the presence of long and thin
absorbing threads along the first flux rope. For event 2, the
filament at the east location of the flux rope may indicate part of
the pre-existing flux rope structures.

When the surge in event 1 appears and the filament in event 2 is
activated, the obvious brightenings and flare activities are
observed simultaneously at the east of the two flux ropes. This
implies that heating takes place and may illuminate the flux rope
body by filling it with hot and dense plasma emitting in the EUV
channels. This is similar to the observations of Raouafi (2009), who
suggested that a C-class flare near one footpoint of the flux rope
led to the brightening of the magnetic structure showing its fine
structure. While the material arrives at the FPs of the fine-scale
structures, the FPs are consequently brightened. The brightening at
the FPs may be caused by the conversion from the kinetic energy to
the thermal energy.

Our observations show that there exist 7$-$15 FPs of the fine-scale
structures for each end of the flux rope. By comparing the EUV
observations with the HMI magnetograms, we find that the FPs at one
end of the flux rope on July 29 are rooted in the same polarity
fields and the FPs at the other end are anchored at the opposite
polarity fields (Figure 3). For the flux rope on August 04, the
eastern end could not be identified and only the western end is
analyzed here. The FPs of the fine-scale structures are located at
network magnetic fields and their magnetic fluxes are in the range
of 5.6$\times$10$^{18}$$-$8.6$\times$10$^{19}$ Mx. The magnetic
fluxes of the two flux ropes are at least 4.3$\times$10$^{20}$ and
7.6$\times$10$^{20}$ Mx. According to the statistical study of Sung
et al. (2009), the magnetic flux of 34 magnetic clouds (MC) for
in-situ observations varies from 1.25$\times$10$^{18}$ to
4.69$\times$10$^{21}$ Mx with the average of 1.1$\times$10$^{21}$Mx.
The magnetic flux of the flux ropes in our observations is
comparable to that of the MC. Moreover, almost half of the FPs of
the fine-scale structures show converging motion of smaller magnetic
structures over 10 hr before the appearance of the flux rope. The
network magnetic field is often thought to be the converging center
(Zhang et al. 1998), which is consistent with our observations. As
small-scale magnetic fields located at the converging centers always
exist tens of hours (Liu et al. 1994), implying that the flux ropes
have a relatively long lifetime.

The flux ropes are observed in all the 7 EUV channels (304, 171,
193, 211, 335, 94 and 131 {\AA}) of the \emph{SDO}/AIA that cover
the temperature from 0.05 MK to 11 MK. This is consistent with
recent observations of Patsourakos et al. (2013) and Li \& Zhang
(2013), who reported the hot and cool components of flux ropes.
However, there exist the flux ropes that could only be observed in
hot channels such as 94 and 131 {\AA} (Zhang et al. 2012; Cheng et
al. 2011, 2012). The comprehensive characteristics of flux ropes
need to be analyzed in further studies.

\acknowledgments {We acknowledge the \emph{SDO}/AIA and HMI for
providing data. This work is supported by the National Basic
Research Program of China under grant 2011CB811403, the National
Natural Science Foundations of China (11025315, 11221063, 10921303
and 11003026), the CAS Project KJCX2-EW-T07, and the Young
Researcher Grant of National Astronomical Observatories, Chinese
Academy of Sciences.}

{}
\clearpage

\begin{figure}
\centering
\includegraphics
[bb=52 98 507 722,clip,angle=0,scale=0.9]{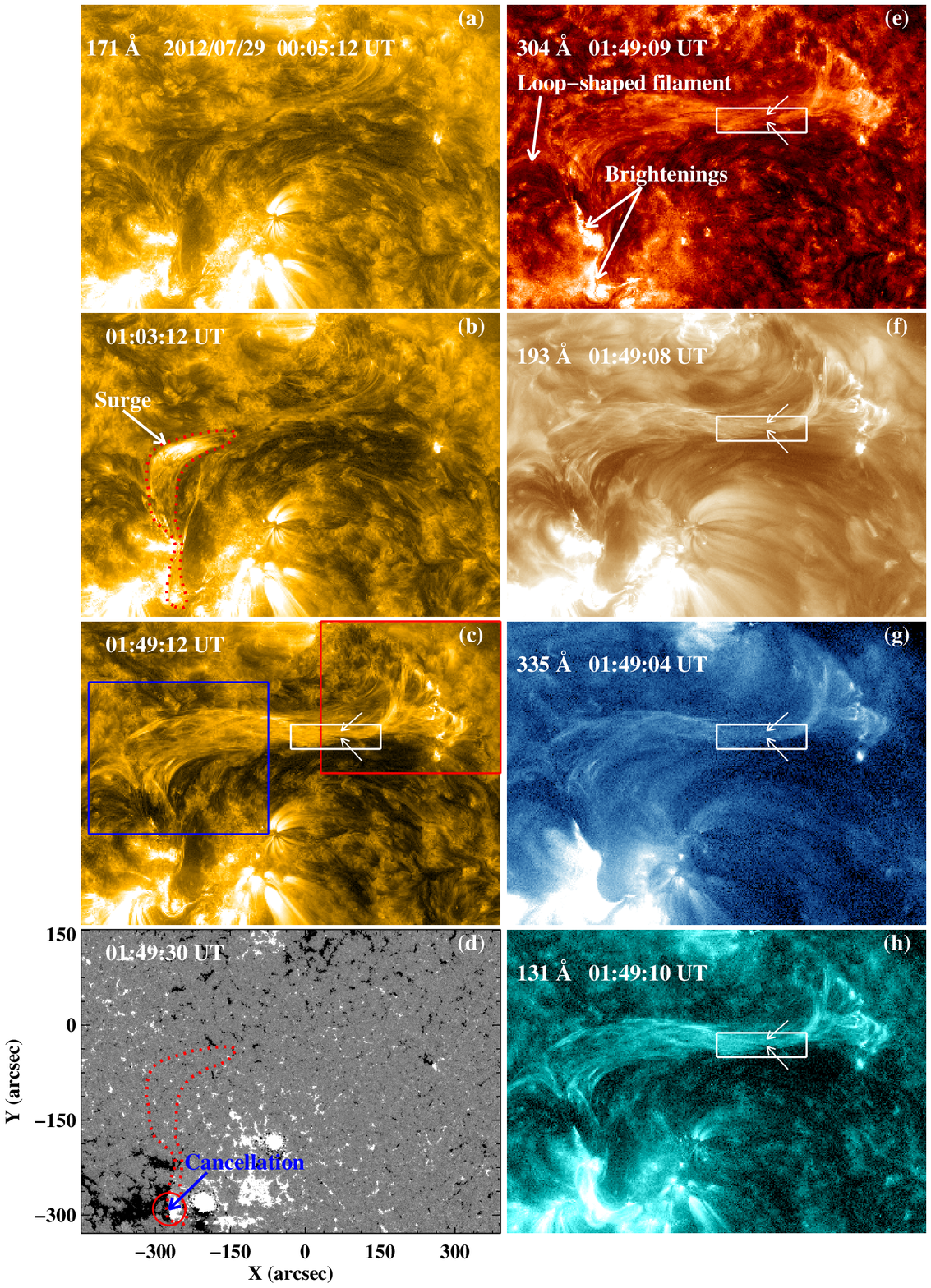} \caption{
{\emph{SDO}}$/$AIA multi-wavelength images and HMI line-of-sight
magnetogram showing the evolution of the flux rope on 2012 July 29
(see Animations 1 and 2, available in the online edition of the
journal). The red solid rectangle in panel \emph{c} denotes the FOV
of Figure 3\emph{a} and the blue one denotes the FOV of Figure
3\emph{d}. \label{fig1}}
\end{figure}
\clearpage

\begin{figure}
\centering
\includegraphics
[bb=52 93 505 729,clip,angle=0,scale=0.9]{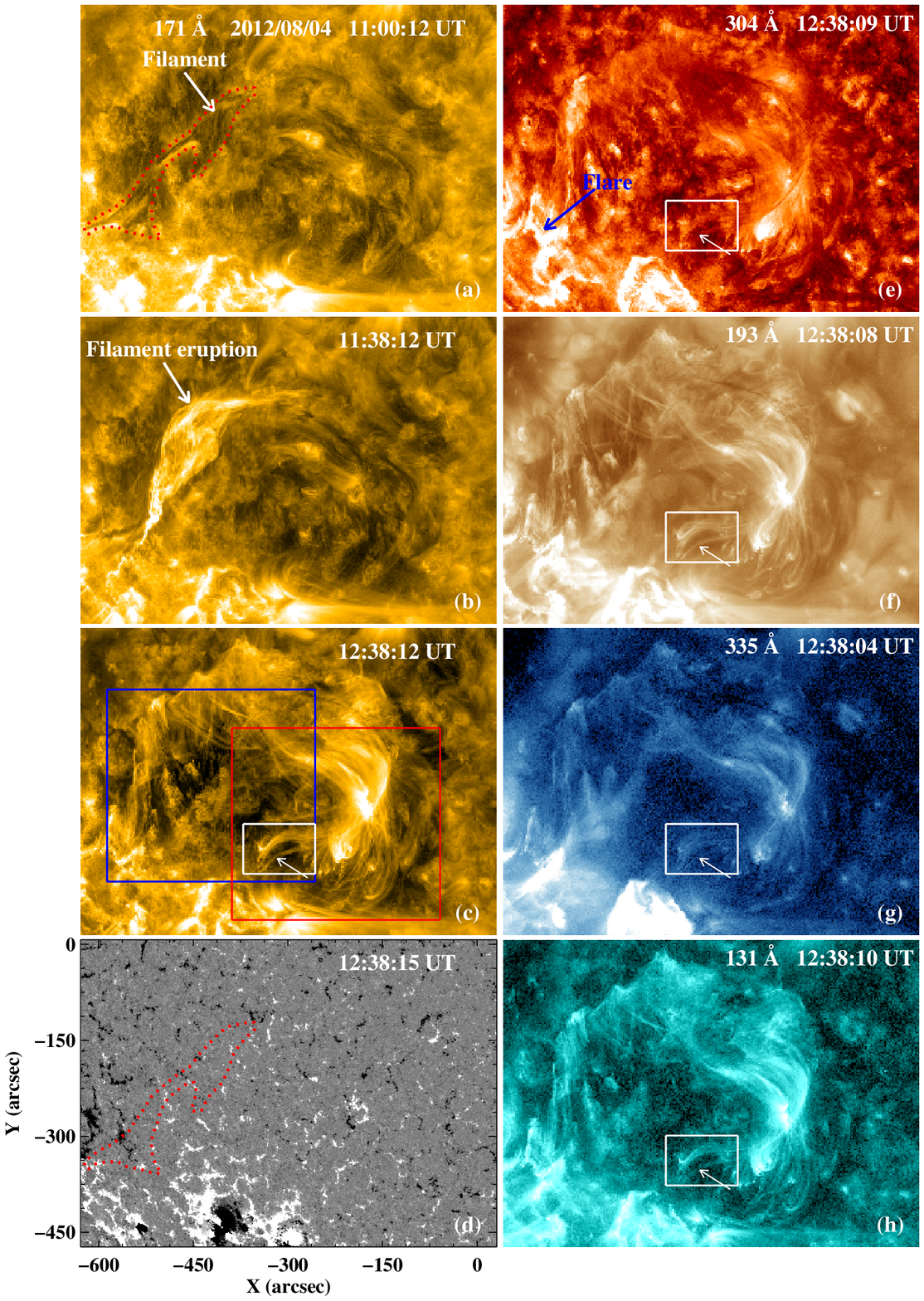}
\caption{{\emph{SDO}}$/$AIA multi-wavelength images and HMI
line-of-sight magnetogram showing the evolution of the flux rope on
2012 August 04 (see Animations 3 and 4, available in the online
edition of the journal). The red solid rectangle in panel \emph{c}
denotes the FOV of Figure 4\emph{a} and the blue one denotes the FOV
of Figure 4\emph{d}. \label{fig2}}
\end{figure}
\clearpage

\begin{figure}
\centering
\includegraphics
[bb=62 127 532 693,clip,angle=0,scale=0.9]{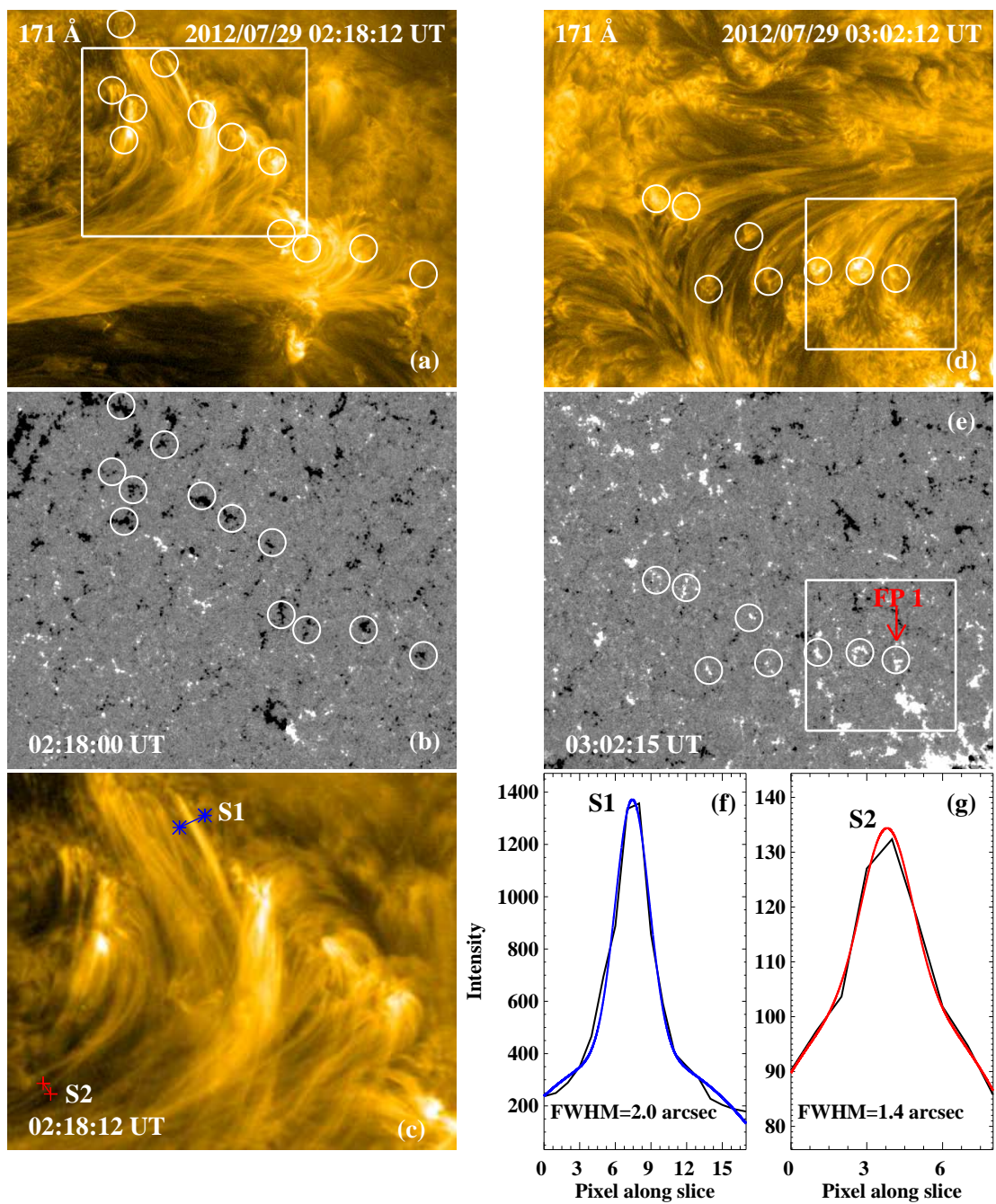}
\caption{{\emph{SDO}}$/$AIA 171 {\AA} images and HMI magnetograms
showing the western (panels \emph{a} and \emph{b}) and eastern ends
(panels \emph{d} and \emph{e}) of the flux rope on 2012 July 29, and
the Gaussian fitting profiles (panels \emph{f} and \emph{g}) showing
the widths of fine-scale structures. The white rectangle in panel
\emph{a} denotes the FOV of panel \emph{c} and white rectangles in
panels \emph{d} and \emph{e} denote the FOV of Figure 5. The blue
and red curves in panels \emph{f} and \emph{g} are respectively the
Gaussian fitting profiles of the intensity-location curves (black
ones) along the blue and red slices (Slices ``S1" and ``S2") in
panel \emph{c}. \label{fig3}}
\end{figure}
\clearpage

\begin{figure}
\centering
\includegraphics
[bb=63 104 530 716,clip,angle=0,scale=0.9]{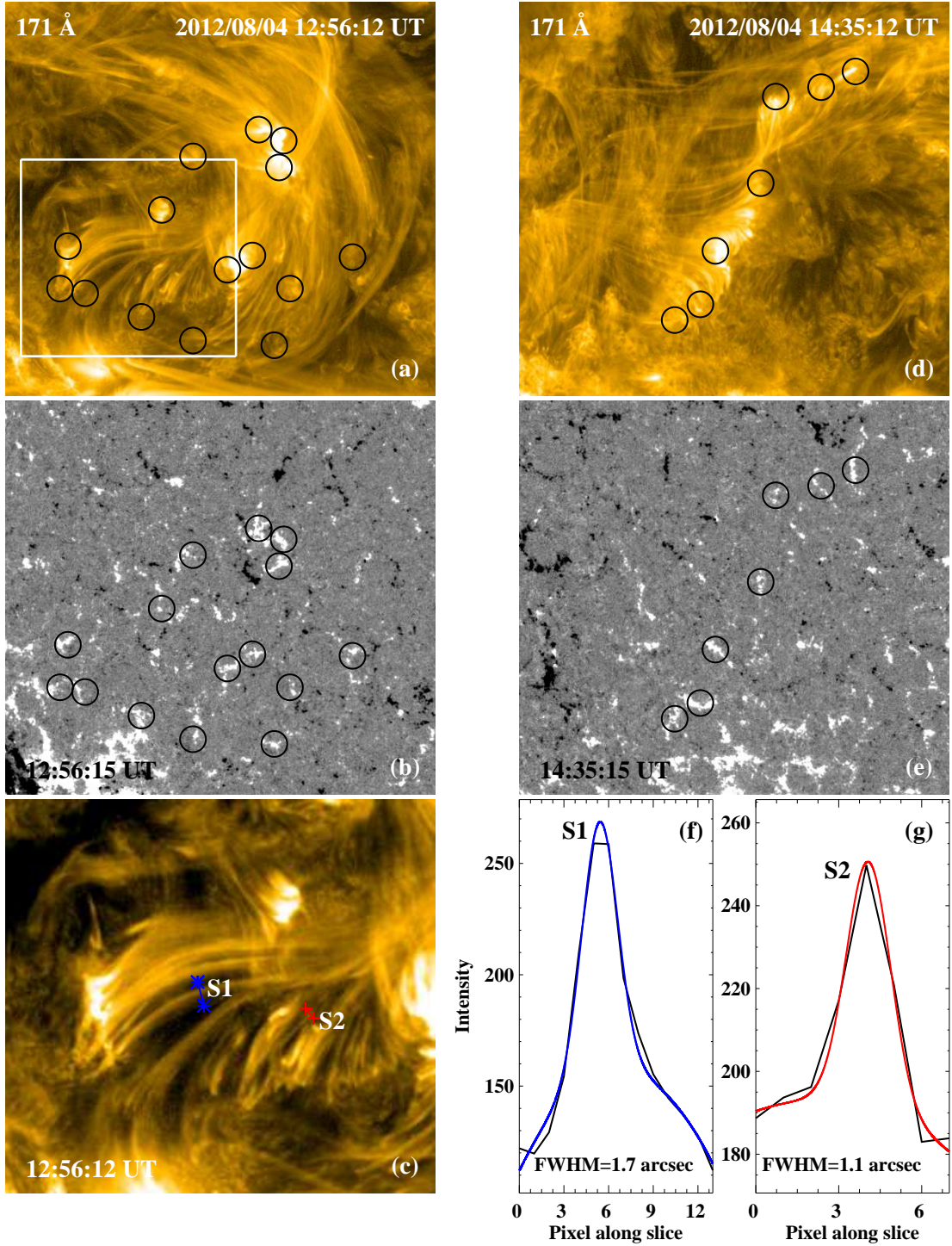}
\caption{{\emph{SDO}}$/$AIA 171 {\AA} images and HMI magnetograms
showing two main western ends of the flux rope (one in panels
\emph{a}$-$\emph{b} and the other in panels \emph{d}$-$\emph{e}) on
2012 August 4, and the Gaussian fitting profiles (panels \emph{f}
and \emph{g}) showing the widths of fine-scale structures. The white
rectangle in panel \emph{a} denotes the FOV of panel
\emph{c}.\label{fig4}}
\end{figure}
\clearpage

\begin{figure}
\centering
\includegraphics
[bb=51 82 510 720,clip,angle=0,scale=0.9]{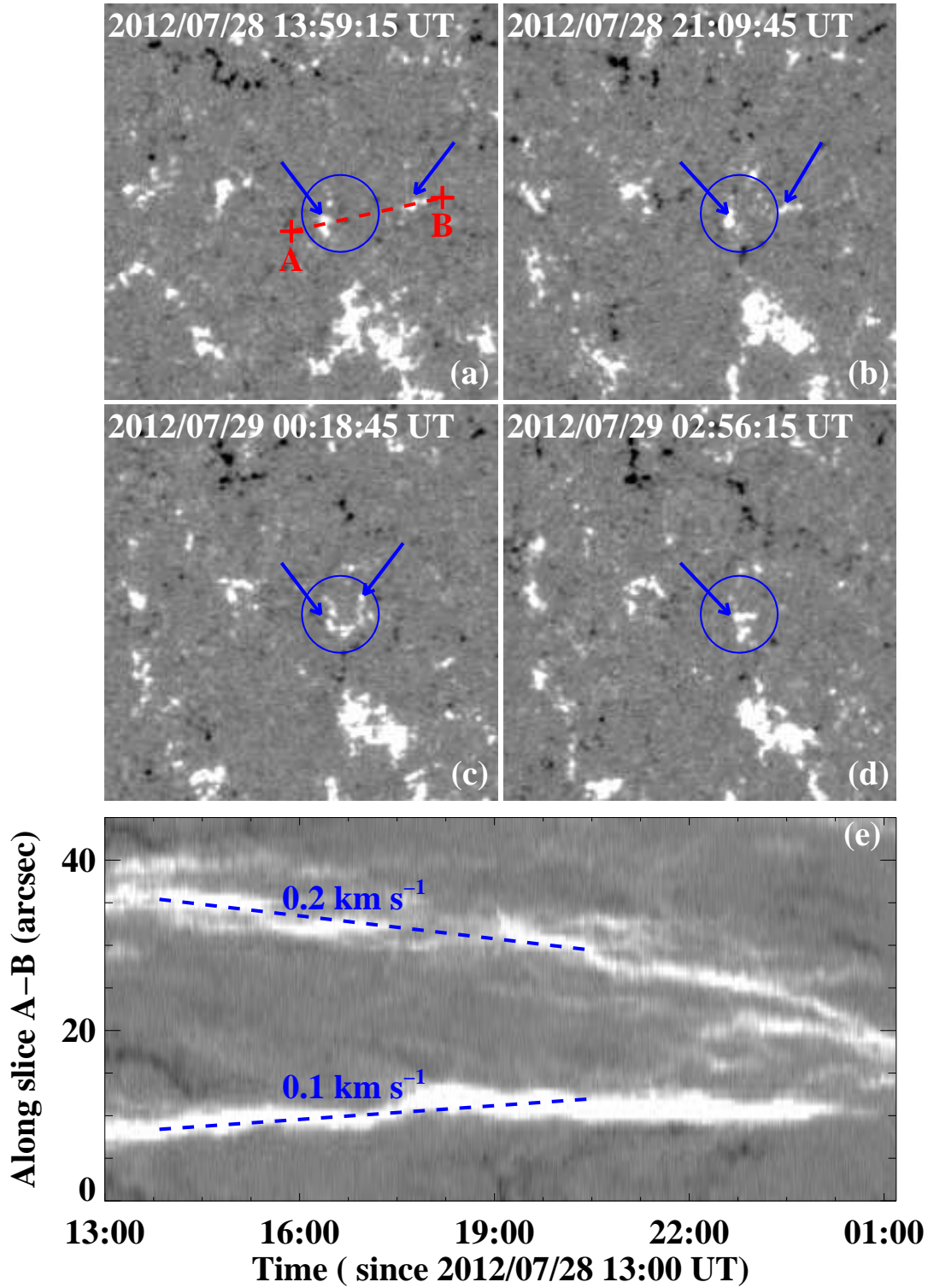}
\caption{{\emph{SDO}}$/$HMI magnetograms showing the converging
motion of smaller magnetic structures at the FPs of the fine-scale
structures. Blue circles denote one of the eastern FPs (``FP 1" in
Figure 3\emph{e}) of the fine-scale structures for the first flux
rope on 2012 July 29. The stack plot along Slice ``A$-$B" (red
dashed line in panel \emph{a}) is shown in panel \emph{e}.
\label{fig5}}
\end{figure}
\clearpage

\end{document}